\documentclass[prb,aps,showpacs,twocolumn]{revtex4}
\usepackage{amssymb}
\usepackage{amsmath}
\usepackage{epsfig}
\usepackage{graphicx}

\begin{document}

\title{First-principles calculations of spin-triplet andreev reflection spectra at half-metallic ferromagnet/superconductor interface}

\author{Ling Tang}

\affiliation{Department of Applied Physics, Zhejiang University of
Technology, Hangzhou 310023, P.~R.~China}

\begin{abstract}
Combining the first-principles noncollinear calculations of
scattering matrices with Andreev approximation, we investigated the
spin-triplet Andreev reflection (AR) spectra for the interface
between half-metallic ferromagnet Co$_{2}$MnSi and \emph{s}-wave BCS
superconductor Al with and without interfacial roughness, where the
orientations of magnetic moments near the interface are randomly
distributed. The calculated results show that the AR spectra have
peak structures near zero bias for the clean interface with relative
weak magnetic disorder. With increasing the degree of interfacial
roughness or magnetic disorder, these subgap peaks of conductance
spectra will be washed out. The results also show that the value of
subgap conductance spectrum can be raised significantly by the
magnetic disorder. Finally, our calculations reveal that the
long-range spin-triplet AR in Co$_{2}$MnSi/Al(001) interface can be
enhanced by a small amount of interfacial roughness.
\end{abstract}

\maketitle

\section{Introduction}

In normal-metal/\emph{s}-wave BCS superconductor heterostructure
(NM/S), below the superconducting energy gap a spin up or down
electron comes into NM/S interface will be reflected as a hole in
the opposite spin subband, which is called spin-singlet Andreev
reflection(AR) process\cite{Andreev-1964} and this phenomenon has
been well studied by the first-principles approach very
recently.\cite{Wang-prb12} If the NM is ferromagnet (F), the number
of propagating states at the Fermi energy is different for majority
and minority spin owing to the presence of exchange splitting.
Therefore only a fraction of incident electron can be Andreev
reflected and then the conductance of this spin-singlet AR is
suppressed significantly.\cite{Beenakker-prl95} The most extreme
case is the half-metallic ferromagnet (HF) with only one spin
subband at Fermi level, thus the spin-singlet AR process will be
totally prohibited in HF/S interface.

However, since the striking observation of Josephson current in
S/HF/S junction has been reported,\cite{Keizer-nature06} the
spin-triplet AR has attracted a great deal of attention in recent
years.\cite{Eschrig-natphy08,Asano-prl07,beri-prb09,Linder-prb10,Valls-prb09,xing-prl07}
It is because that the exotic phenomena such as 0-$\pi$ transitions
in S/HF/S Josepson junction\cite{Volkov-prl03,Bergeret-rmp05} may
have some potentially applications on superconducting quantum
computing circuits. At HF/S interface, the spin-triplet AR means
that the incident electrons undergo a spin-flip scattering and
penetrate into BCS superconductor as Cooper pair, leaving a phase
coherence hole in the same spin which can be survived at HF side. So
the incident electrons and phase-coherently reflected holes (or vice
verse) within the same spin subband can induce the spin-triplet
superconducting correlation in HF. Further, these spin-triplet
pairing is immune to the exchange splitting in the ferromagnet,
which results in the long-range proximity
effect\cite{Bergeret-rmp05} as well as the supercurrent across the
S/HF/S heterostructure. In addition to all this, the spin-triplet AR
induced by spin-flip scattering at HF/S also has been proposed to
explain the experimental spin polarization (P) of HF deviated from
its theoretical prediction P=100\%.\cite{Eschrig-prl10}

Up to now, the theoretical calculations about spin-triplet AR
generally deal with the interface properties by a simple
delta-function potential
barrier\cite{Kupferschmidt-prb09,Brouwer-prb11,Brouwer-prb12,Brouwer-prb11b}
or scattering matrices coefficients with model
parameters,\cite{Eschrig-prb10} where the complex Fermi surfaces and
electronic band structure of real materials have not been considered
yet. For realistic HF/S interface, it is argued that multichannel
scattering theory of propagating states at Fermi surface should be
taken into account for calculating AR conductance.\cite{xia-prl02}
Moreover, the symmetry of band states when matching wave-function
between two bulk crystal has effect on transport properties of the
interface, which can be large and should not be
neglected.\cite{xia-prb01} Although the first-principles method has
been applied to AR of Fe/Al interface with
spin-flip,\cite{shuai-prb10} it is difficult to distinguish the
spin-singlet and spin-triplet AR in normal F (with incomplete spin
polarization), while in full HF there is only spin-triplet AR.
Therefore, the main purpose of this paper is to perform the
first-principles density functional theory calculation without
introducing any arbitrary parameters to study the spin-triplet AR
conductance spectra in HF/S.

On the other hand, the well-known Heusler alloy Co$_{2}$MnSi belongs
to HF and can be grown by sputtering techniques.\cite{heusler} It
also can be well matched in the (001) direction with \emph{s}-wave
BCS superconductor Al by rotating $45^{\circ}$, which makes the
Co$_{2}$MnSi/Al(001) interface a good candidate for studying
spin-triplet AR in details.

In this paper, we use the method which combined the scattering
matrix approach\cite{shuai-prb08,xia-prb06,tang-mplb08} with Andreev
approximation\cite{Andreev-approx} in the frame of Bogoliubov-de
Gennes equations\cite{Bogoliubov} to evaluate the spin-triplet AR
conductance in Co$_{2}$MnSi/Al(001) interface with and without
interfacial roughness. The spin-flip scattering process at HF/S
required by spin-triplet AR is introduced by the disordered
distribution of interfacial magnetic moments direction. Here the
transmission and reflection matrices of magnetic noncollinear
textured HF/S interface are calculated by first-principles
scattering wave-function matching method.\cite{shuai-prb08}

Our results of AR conductance spectra show that for clean interface
without roughness the relative weak disorder of interfacial magnetic
moments can induced a conductance peak near zero bias. However, this
subgap conductance peak structure will vanish as the degree of
interfacial roughness or magnetic disorder increasing. In addition,
our calculations also show that with increasing magnetic disorder
the value of AR conductance spectrum can be remarkablely raised.
Finally, our results demonstrate that a small amount of interfacial
roughness can enhance the spin-triplet AR process.

\section{Method}

\begin{figure}
  \includegraphics[width=8.6cm, bb=35 15 542 412]{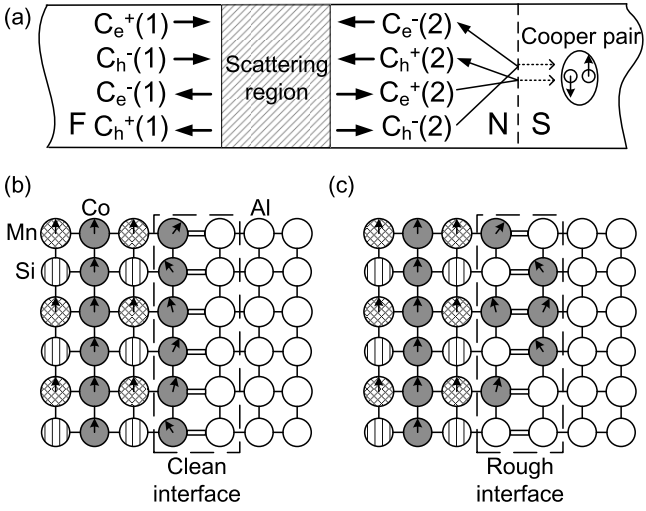}\\
  \caption{(a) The sketch of the whole scattering
problem. Region N is a fictitious layer of superconductor in its
normal states. At the N/S interface, there is only AR process which
connected the incident electron with reflected hole or vice versa
leaving a Cooper pair in S. (b) Clean and (c) rough
Co$_{2}$MnSi/Al(001) interface model with CoCo/Al termination in
scattering region respectively. The small arrows indicate the
direction of magnetic moments and the disordered distribution of
moments is only within the interfacial layers.
  }\label{f1}
\end{figure}

In order to investigate AR within scattering matrix framework,
following Beenakker's model\cite{Andreev-approx} we consider a
general F/S junction with inserting a fictitious region N between F
and S, which is composed of S in its normal state. The F/N interface
with ideal F as left lead and ideal N as right lead constitute a
well-defined scattering problem. Meanwhile, there is only AR at N/S
interface owing to no potential difference at this interface. The
sketch of the whole scattering problem including AR is shown in Fig.
\ref{f1}(a).

At the F/N interface, the incoming and outgoing scattering state
coefficients for electrons and holes quasiparticles at excitation
energy $\varepsilon$ can be written as\cite{shuai-prb10,lambert}
\begin{equation}
\left(
\begin{array}
[c]{c}%
C_{e}^{-}\left(  1\right) \\
C_{e}^{+}\left(  2\right) \\
C_{h}^{+}\left(  1\right) \\
C_{h}^{-}\left(  2\right)
\end{array}
\right)  =\left(
\begin{array}
[c]{cccc}%
r_{11}^{e}(\varepsilon) & t_{12}^{e}(\varepsilon) & 0 & 0\\
t_{21}^{e}(\varepsilon) & r_{22}^{e}(\varepsilon) & 0 & 0\\
0 & 0 & r_{11}^{h}(\varepsilon) & t_{12}^{h}(\varepsilon)\\
0 & 0 & t_{21}^{h}(\varepsilon) & r_{22}^{h}(\varepsilon)%
\end{array}
\right)  \left(
\begin{array}
[c]{c}%
C_{e}^{+}\left(  1\right) \\
C_{e}^{-}\left(  2\right) \\
C_{h}^{-}\left(  1\right) \\
C_{h}^{+}\left(  2\right)
\end{array}
\right)\label{eq1}%
\end{equation}
where subscript 1 refers to F, 2 to N and $+$ ($-$) denotes right or
left going waves as shown in Fig. \ref{f1}(a). The coefficient
vectors $C_{e(h)}^{\pm}(i)$ ($i=1,2$) are amplitudes of propagating
Bloch states of electron (hole) in the left lead F and right lead N.
Here $r_{ij}^{e(h)}(\varepsilon)$ and $t_{ij}^{e(h)}(\varepsilon)$
($i,j=1,2$) is reflection and transmission matrix of electron (hole)
for F/N interface which is $2\times2$ block matrix in spin space. In
our calculation the spin-flip scattering process originating from
noncollinear magnetic moments distribution at F/N interface is just
described by the off-diagonal scattering matrix elements in this
spin space.

Moreover, these normal-state reflection and transmission matrix can
be calculated by a two-steps procedure. First the effective
single-electron potential for collinear F/N interface that obtained
by density functional theory will serve as input to the second step.
Secondly, in the rigid potential approximation the noncollinear
transport coefficients can be evaluated using wave-function matching
method.\cite{shuai-prb08}

Considering the right and left going quasiparticles in N region is
connected by AR at N/S interface, the reflection matrix for the
whole F/N/S system can be obtained by solving the above scattering
equation Eq. (\ref{eq1}). According to Beenakker's
model,\cite{Andreev-approx} we introduce Andreev approximation at
N/S interface, i.e., $C_{e}^{-}(1)=\alpha C_{h}^{-}(1)e^{i\phi}$ and
$C_{h}^{+}(1)=\alpha C_{e}^{+}(1)e^{-i\phi}$, where $\phi$ is phase
of superconductor. The factor
$\alpha=\exp[-i\arccos(\varepsilon/\Delta_{0})]$ for
$|\varepsilon|<\Delta_{0}$ and
$\alpha=[\varepsilon-sgn(\varepsilon)\sqrt{\varepsilon^{2}-\Delta_{0}^{2}}]/\Delta_{0}$
for $|\varepsilon|>\Delta_{0}$, where $\Delta_{0}$ is \emph{s}-wave
superconductor pair potential at zero temperature. Therefore
substituting the formula of Andreev approximation into Eq.
(\ref{eq1}), we can obtain the relationship of the incoming and
reflected states only at F region, which determines the reflection
matrix of whole F/N/S system. Namely,
\begin{align}
R_{ee} &  =r_{11}^{e}(\varepsilon)+\alpha^{2}t_{12}^{e}(\varepsilon)r_{22}^{h}(\varepsilon)\left[  1-\alpha^{2}%
r_{22}^{e}(\varepsilon)r_{22}^{h}(\varepsilon)\right]  ^{-1}t_{21}^{e}(\varepsilon)\label{eq2}\\
R_{he} &  =\alpha e^{-i\phi}t_{12}^{h}(\varepsilon)\left[  1-\alpha^{2}r_{22}^{e}(\varepsilon)r_{22}%
^{h}(\varepsilon)\right]  ^{-1}t_{21}^{e}(\varepsilon)\label{eq3}%
\end{align}
So in the linear-response regime, the total conductance of F/S is
\begin{equation}\label{eq4}
    G_{\mathrm{FS}}\left(  \varepsilon\right)  =\frac{e^{2}}{h}\text{Tr}\left(1-R_{ee}R_{ee}^{\dagger}+R_{he}R_{he}^{\dagger}\right)%
\end{equation}
and we define the spin-triplet AR conductance is
\begin{equation}
G_{\mathrm{triplet}}(\varepsilon)\equiv\frac{e^{2}}{h} \text{Tr}(
R_{he}^{\downarrow \downarrow}R_{he}^{\downarrow\downarrow\dagger} +
R_{he}^{\uparrow\uparrow}R_{he}^{\uparrow\uparrow\dagger})
\label{eq5}%
\end{equation}
In order to study the character of subgap AR conductance spectrum,
here we define the normalized conductance
$g(V)\equiv[G_{\mathrm{FS}}(V)-G_{\mathrm{FN}}(0)]/G_{\mathrm{FN}}(0)$,
where $G_{\mathrm{FN}}(0)$ is conductance when the superconductor in
its normal-state and $V$ is normalized bias defined as $V\equiv
\varepsilon/\Delta_{0}$. Here the zero temperature pair potential is
very small, e.g., $\Delta_{0}(\mathrm{Al})\approx0.34$ meV.\cite{al}
The difference of electronic structure within this tiny energy
region can be ignored, so in our calculations the reflection and
transmission matrices at excitation energy $\varepsilon$ in Eq.
(\ref{eq2}) and (\ref{eq3}) is approximately equal to the transport
coefficients at Fermi level.

\section{Computational Details}

Co$_{2}$MnSi is a representative compound of Heusler alloys, where
the Co atoms form a simple cubic lattice while the Mn and Si atoms
are situated the alternate body center positions. The experimental
lattice constant of bulk Co$_{2}$MnSi is 10.68 [a.u.]\cite{lattice}
and the \emph{s}-wave superconductor Al has fcc crystal structure
with lattice constant 7.653 [a.u.].\cite{al} So the fcc Al and
Co$_{2}$MnSi lattice can be matched in (001) direction, by rotating
$45^{\circ}$ around (001) axis for fcc Al lattice. The mismatch of
this heterostructure is about 1.3\% and we keep the lattice constant
of bulk Co$_{2}$MnSi unchanged and slightly compress the bulk Al
lattice.

In our calculations, the atom sphere approximation (ASA)\cite{asa}
is taken into account for obtaining electronic structure and
conductance. These radius of spheres are determined by the crystal
Hartree potential\cite{sphere} with filling up all the space. The
atom spheres of the Co$_{2}$MnSi are chosen as
$r(\mathrm{Co})=2.576$ [a.u.] and
$r(\mathrm{Mn})=r(\mathrm{Si})=2.681$ [a.u.]. Similarly, the radius
of Al atom sphere is chosen as $r(\mathrm{Al})=2.951$ [a.u.]. Due to
the CoCo monolayer and MnSi monolayer stack alternately in the (001)
direction, the Co$_{2}$MnSi(001) has two kinds of terminations (CoCo
and MnSi) adjacent to Al lead. The distance of these two ideal
CoCo/Al and MnSi/Al interfaces can be determined as follows. From
Fig. \ref{f1}(b), one can see that the interfacial cubic unit cell
consists of 0.5 Co (0.25Mn, 0.25Si) and 0.5 Al, where these Co
(MnSi) and Al should fill up the space of the cubic with volume
$V=d\times(a_{\mathrm{Co_{2}MnSi}}/2)^{2}$. Since we have chosen the
radius of Co (MnSi) and Al, the interface distance between CoCo
(MnSi) and Al monolayer $d$ can be determined by the above formula.
In addition, the other distance for Al-Al and Co-MnSi layers still
maintain their bulk values.

The structural model for alloyed rough interface of CoCo/Al and
MnSi/Al are constructed by a $5\times5$ lateral supercell which
contains 50 interfacial Co atoms in CoCo/Al and 25 interfacial Mn
atoms in MnSi/Al, where the lattice has no distortion but the
interfacial atoms occupy the sites randomly and the possibility of
occurrence is according to their concentrations $x$.  As shown in
Fig. \ref{f1}(c), here we assume that Co (MnSi) and Al atoms diffuse
into each other for one monolayer at interface, which can be denoted
as [Co$_{1-x}$Al$_{x}|$Co$_{x}$Al$_{1-x}$] or
[(MnSi)$_{1-x}$Al$_{x}|$(MnSi)$_{x}$Al$_{1-x}$]. For MnSi/Al
interface we also assumed that the Mn and Si have the same chance to
diffuse into Al monolayer. Furthermore, in our rough interface model
the numbers of interfacial Co (MnSi) and Al atoms are equal to that
of clean interface. Hence, for filling the interfacial space, we
take the distance between CoCo (MnSi) and Al monolayer of rough
interface as same as the values of clean interface.

In this paper, the effective single-electron potential of
Co$_{2}$MnSi/Al(001) interface is obtained from tight-binding
linearized muffin-tin-orbital(TB-LMTO) surface Green function method
with ASA approximation.\cite{SGF,Andersen-prl84,Andersen-prb86} For
rough interface, the coherent potential approximation
(CPA)\cite{SGF,Velicky-pr69,ke-prl08} is used to deal with the
alloyed interfacial roughness. After performing self-consistent
electronic structure calculation, we put the atomic sphere
potentials into the corresponding sites for evaluating the AR
conductance. Here the exchange-correlation potential is taken within
local spin density approximation (LSDA) using von Barth-Hedin
parametrization.\cite{LSDA} The electron is treated
scalar-relativistically and the cutoff of orbital angular momentum
of basis is $l_{\mathrm{max}}=2$, corresponding to \emph{spd}-basis.
The reciprocal lattice vectors of Brillouin zone (BZ) are each
divided into NK=12 intervals for calculating the self-consistent
electronic structure of interface. Meanwhile, the calculations of AR
conductance are performed with a $k_{\|}$-mesh density equivalent to
2500 $k_{\|}$-mesh points in the 2D BZ of $1\times1$ lateral unit
cell.

Here we assume that the spin-flip scattering, which is necessary for
spin-triplet AR in full HF, only originates from the noncollinear
random distribution of interfacial magnetic moments direction.
Therefore the magnetic disorder in interfacial magnetic atoms is
also simulated by $5\times5$ lateral supercell as the structural
model for rough interface. For these interfacial magnetic atoms, the
deviation angle between the local magnetic moment direction and the
global quantum axis satisfy Gaussian random distribution, where the
distribution width is $\Delta\theta$. In addition, the rigid
potential approximation\cite{shuai-prb08} has been employed here and
the spin-orbit coupling is neglected in our noncollinear
calculations.

\section{Results and Discussion}

\begin{figure}
  \includegraphics[width=8.6cm]{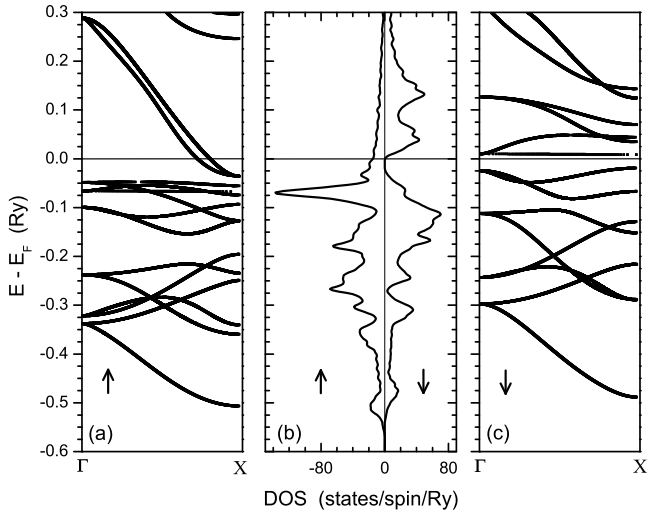}\\
  \caption{The electronic structure of bulk Co$_{2}$MnSi. (a) and (c)
The energy band along $\Gamma$-X direction for majority and minority
channel respectively. (b) The spin-resolved density of states (per
formula unit).
  }\label{f2}
\end{figure}

\begin{figure}
  \includegraphics[width=8.6cm]{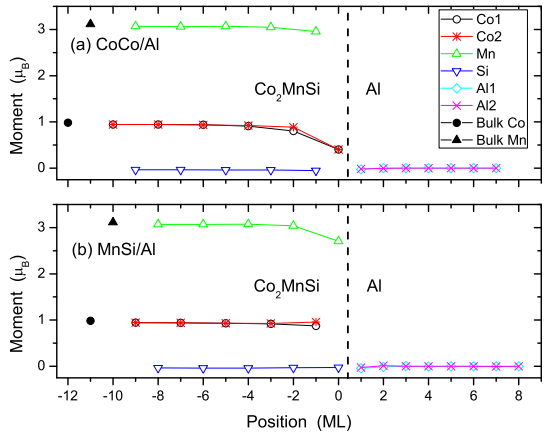}\\
  \caption{The layer-resolved magnetic moments for (a) CoCo- and (b)
MnSi- terminated Co$_{2}$MnSi/Al(001) interfaces without interfacial
roughness. The magnetic moments of Co or Mn atoms have been
suppressed at the interface and recover to their bulk values away
about 4 [ML] from interface.
  }\label{f3}
\end{figure}

In this section, we first examine the self-consistent calculated
electronic structure of bulk Co$_{2}$MnSi as left lead and
Co$_{2}$MnSi/Al(001) interface respectively. The resulting band
structure along $\Gamma$-X direction and density of states (DOS) for
bulk Co$_{2}$MnSi are shown in Fig. \ref{f2}. The shape of DOS
agrees well with the previous results reported in Ref.\cite{dos}.
The majority band is totally metallic and there is a semiconductor
energy gap ($\sim$ 0.4 eV in agreement with Ref.\cite{gap}) in
minority spin. The Fermi level is inside the minority spin gap near
the conduction band minimum, which indicates that the bulk
Co$_{2}$MnSi is a full half-metallic ferromagnet and the current
injected into Co$_{2}$MnSi/Al(001) interface from left lead has
100\% spin polarization. Further, Fig. \ref{f3} shows the
layer-resolved magnetic moments for CoCo- and MnSi-terminated
Co$_{2}$MnSi/Al(001) interfaces without interfacial roughness. It
can be seen that the interfacial magnetic moments are suppressed due
to the charge transfer between Co (Mn) and Al atoms. However, the
magnetic moments of Co and Mn atoms can recover to their bulk values
just about 4 [ML] away from interface. In addition, the results for
rough Co$_{2}$MnSi/Al(001) interface is similar to that of clean
interface and have not been demonstrated here for simplicity.

\begin{figure}
  \includegraphics[width=8.6cm, bb=0 0 485 372]{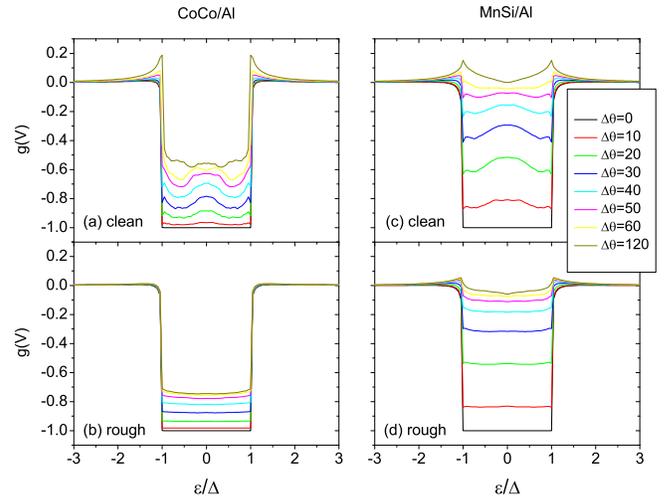}\\
  \caption{The normalized conductance
spectra for different degree of interfacial magnetic disorder
($\Delta\theta$). The interfacial alloy concentration of rough
interface is 50\%-50\% ($x=0.5$). For clean interface with
$\Delta\theta<60^{\circ}$, it is clear that the peak structure
emerges near the zero bias. However, the larger amount of magnetic
disorder ($\Delta\theta>60^{\circ}$) or interfacial roughness can
wash out this subgap conductance structure.
  }\label{f4}
\end{figure}

Next we concentrate on the conductance spectra of
Co$_{2}$MnSi/Al(001) interface. Fig. \ref{f4} shows the normalized
conductance spectra of Co$_{2}$MnSi/Al(001) interface for different
$\Delta\theta$. In the case of $\Delta\theta=0^{\circ}$ the
conductance below the gap is exactly zero because the left lead
Co$_{2}$MnSi only has majority spin band at the Fermi energy and the
AR process is prohibited without spin-flip. Once there is spin-flip
scattering created by the noncollinear random distribution of
interfacial magnetic moments direction, a conductance peak structure
will emerge gradually at zero bias as shown in Fig. \ref{f4}(a) and
(c). This zero bias conductance peak (ZBCP) has the largest relative
height when $\Delta\theta\approx30^{\circ}$ and will convert into a
small zero bias conductance dip (ZBCD) with increasing
$\Delta\theta$. The presence of ZBCP in our calculation can be
attributed to the Fermi wave-vector mismatch between Co$_{2}$MnSi
and Al combined with high spin polarization in ferromagnet, which
has been predicted by the extend BTK theory in
Ref.\cite{Valls-prb00}. For $\Delta\theta>60^{\circ}$ the
above-mentioned ZBCP or ZBCD structure have both vanished and the
subgap conductance becomes V shape. Moreover, Fig. \ref{f4}(b) and
(d) shows the conductance spectra of rough Co$_{2}$MnSi/Al(001) with
50\%-50\% interfacial alloy concentration ($x=0.5$). One can observe
that the interfacial alloy roughness can also totally wash out the
ZBCP structure for all $\Delta\theta$. Considering that these subgap
conductance resonance peaks is sensitive to the phase of reflection
coefficients,\cite{Cottet-prb05,Cottet-prb08} so the disappearance
of peak structure can be attributed to the random phase shift gained
at Co$_{2}$MnSi/Al(001) interface.

Now turning our attention to the value of normalized conductance
below the gap, we can see that it increases with increasing
$\Delta\theta$ as shown in Fig. \ref{f4}. In general, the extend BTK
theory\cite{BTK,extend-BTK2,extend-BTK3} is employed to fit this
conductance spectrum of F/S with parameters including barrier
strength Z, zero temperature gap $\Delta_{0}$ and the spin
polarization (P) of ferromagnet which is an important parameter in
spintronics. However, although the ferromagnet is full half-metallic
predicted by theories, the extracted spin polarization from point
contact spectrum of HF/S will be less than 100\% as long as the
conductance around zero bias is not precisely
zero.\cite{Eschrig-prl10} For example, it is found that the subgap
conductance is nonzero in many point-contact AR experiments of HF/S
interface\cite{PCAR1,PCAR2,PCAR3,PCAR4,PCAR5}, which lead to the
obtained P value ranging from 0.5 to 1. But introducing the
spin-mixing parameter associated with spin-flip,\cite{Eschrig-prl10}
one can well fit the conductance spectrum with P=100\%. Similarly,
our first-principles calculated results of Co$_{2}$MnSi/Al(001)
interface show that the spin-flip induced by random distribution of
interfacial magnetic moments direction also can reduce the extracted
P value from spectrum, while the spin polarization of Co$_{2}$MnSi
at Fermi level is still 100\%.

\begin{figure}
  \includegraphics[width=8.6cm, bb=0 0 482 375]{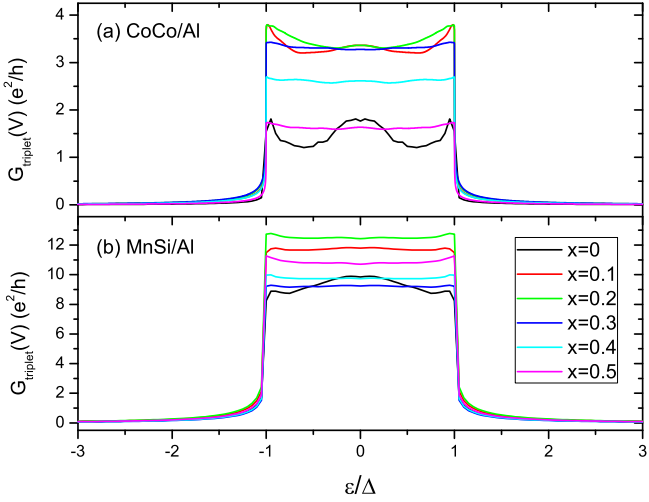}\\
  \caption{The spin-triplet AR
conductance spectra with $\Delta\theta=30^{\circ}$ for different
$x$. The conductance peak structure will vanish for the case of
$x>0.2$ due to the random of phase shift gained at interface.
  }\label{f5}
\end{figure}

Here let us take rough estimation of P using
formula\cite{Soulen-sci}
$G_{\mathrm{FS}}(0)/G_{\mathrm{FN}}(0)=2(1-P)$, where
$G_{\mathrm{FN}}(0)$ is the normal-state conductance and
$G_{\mathrm{FS}}(0)$ is AR conductance at zero bias. For rough
Co$_{2}$MnSi/Al(001) with $\Delta\theta<60^{\circ}$, as seen in Fig.
\ref{f4}(b) and (d), the extracted P ranges from 0.88 to 1 for
CoCo/Al and 0.54 to 1 for MnSi/Al respectively. Comparing between
the CoCo- and MnSi-terminated interface, we see that the subgap
normalized conductance as well as the estimated P of MnSi/Al is much
larger than that of CoCo/Al, which is a consequence of the relative
stronger spin-flip scattering induced by magnetic disorder at
MnSi/Al. This is because that the magnetic moments of interfacial Mn
atoms is larger than that of Co atoms and for the same deviation
angle the atoms with lager magnetic moments can generate stronger
spin-flip scattering. Furthermore, although it is lack of detailed
information about the magnetic disorder in Co$_{2}$MnSi/Al(001), the
distribution width $\Delta\theta$ in ferromagnet/nonmagnetic metal
interface can be up to $40\sim70^{\circ}$ according to
first-principles calculations.\cite{tang-mplb08,Oparin-jap99}
Therefore, our results imply that the interfacial magnetic disorder
in point contact AR experiments can explain the extracted P deviated
from theoretical value P=100\%.

\begin{figure}
  \includegraphics[width=8.6cm, bb=0 0 488 375]{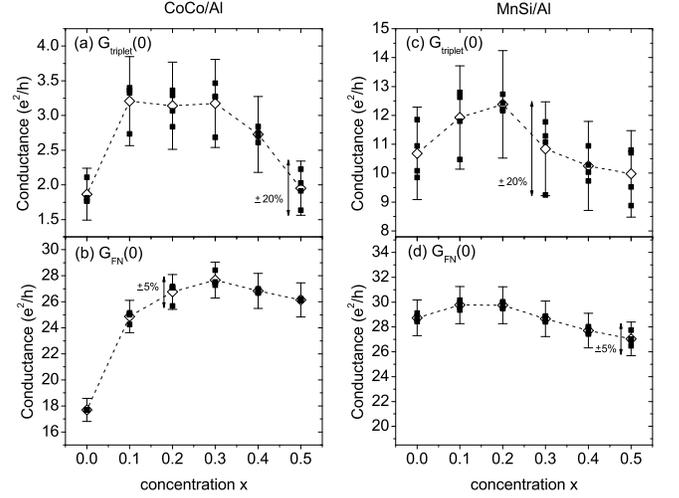}\\
  \caption{The spin-triplet AR and normal-state
conductance at zero bias as function of alloy concentration $x$. The
hollow diamonds denote the average conductance values of different
disorder configurations. The error bars indicate $\pm20\%$ of
average values in (a) and (c) and $\pm5\%$ in (b) and (d). One can
see that the spin-triplet AR has been enhanced by interfacial alloy
roughness around $x=0.2$, which is mainly owing to the increasing of
transparency at interface.
  }\label{f6}
\end{figure}

Finally, in order to study the dependence of spin-triplet AR on
interfacial alloy concentration, we have calculated the spin-triplet
AR using Eq. (\ref{eq5}) with concentration from $x=0.1$ to 0.5.
Fig. \ref{f5} shows the spin-triplet AR conductance spectra with
$\Delta\theta=30^{\circ}$ for different $x$. When the concentration
$x<0.2$ the AR spectrum is similar to that of clean interface
($x=0$) and once $x$ is greater than 0.2 the conductance peak
structure disappeared owing to the random of phase shift at
Co$_{2}$MnSi/Al(001) interface.

Fig. \ref{f6} shows the zero bias spin-triplet AR conductance
$G_{\mathrm{triplet}}(0)$ and normal-state conductance
$G_{\mathrm{FN}}(0)$ at zero bias as function of $x$. In order to
check the statistical error, the results given for different
randomly generated disorder configuration are also shown in Fig.
\ref{f6}. One can see that the variation of sample to sample is
larger for spin-triplet AR ($\sim\pm20\%$) than for normal-state
conductance ($<\pm5\%$) due to the AR is more sensitive to the
fluctuation of interfacial phase shift. It is noted that the
expression of AR coefficient in Eq. (\ref{eq3}) has the form like
multiple beam interference, suggesting the AR conductance is more
sensitive to the amplitude of normal-state scattering coefficients
and phase coherence of the interface. As a consequence, the spread
of spin-triplet AR is more pronounced than that of normal-state
conductance.

Here as a coarse qualitative examination, next we only consider the
average values of the calculated conductances. It can be seen that
as the interfacial Al and Co (MnSi) atoms mixing with each other,
the spin-triplet AR will be enhanced and reach maximum around
$x\approx0.2$. When the interfacial alloy concentration $x$
increases up to 0.5, the spin-triplet AR will go back to be around
the value of clean interface. Further, in low concentration $x$ the
interfacial alloyed atoms can be regard as the impurity at
interface, such as Al atoms in CoCo (MnSi) monolayer. Hence, our
results of AR conductance in Co$_{2}$MnSi/Al(001) demonstrate that
the impurity at interface can assist the spin-triplet AR process
based on first-principles calculations, which agrees with the
prediction in the recent theoretic investigation by model
Hamiltonian.\cite{Brouwer-prb12}

From Eq. (\ref{eq3}) and (\ref{eq5}) one can see that the
spin-triplet AR conductance is evaluated by normal-state
transmission and reflection coefficients, so the dependence of
spin-triplet AR on $x$ is mainly dominated by the transparency of
interface. As shown in Fig. \ref{f6}, the normal-state and
spin-triplet AR conductance has the similar tendency on $x$, but the
relative range of variation is larger for spin-triplet AR. With
increasing $x$, the effect of interfacial alloy can smooth the
potential step within the two layers adjacent to interface, hence
the electronic structure mismatch between two leads is reduced,
which results in lower normal-state reflection and
resistance.\cite{xia-prb06} For example, in Fe/Cr interface it has
been found the interfacial roughness can raise the conductance by
three times.\cite{xia-prb01} On the other hand, with increasing the
roughness the phase coherence of transport process will be
destroyed, which generally suppress the conductance. So combining
the two above-mentioned effects on AR conductance, it is reasonable
that the spin-triplet AR will reach maximum at some interfacial
alloy concentration $x$.

\section{Summary}

In summary, in order to study the effect of interfacial magnetic
disorder on spin-triplet AR spectra, we calculated the scattering
matrices of Co$_{2}$MnSi/Al(001) interface in normal-state by
first-principles noncollinear transport calculation. Next, combining
scattering theory with Andreev approximation, we obtained the
spin-triplet AR spectra for clean and rough interface with different
degree of interfacial magnetic disorder.

The calculated results show that there is a conductance peak below
the superconducting energy gap near zero bias for clean interface
with interfacial magnetic disorder. Once the fluctuation of magnetic
moments directions is sufficient strong or the interface has a small
amount of roughness, this subgap conductance peak structure will be
smeared out. In addition, our calculations show that the value of
subgap conductance will increase with increasing the interfacial
magnetic disorder. Therefore we argue that the reduction of
experimental spin polarization from observation of nonzero subgap
conductance in HF is owing to interfacial magnetic disorder effect.
Meanwhile, the spin polarization of bulk HF is still 100\%.
Moreover, we also found that for rough interface the spin-triplet AR
conductance at zero bias has the maximum value with interfacial
alloy concentration around $x\approx0.2$, which suggests that the
impurity at interface can enhance the spin-triplet AR process.

\section*{Acknowledgments}

The authors acknowledge Prof. Ke Xia for suggesting the problem and
Dr. Shuai Wang for useful discussion about the calculations. We are
also grateful to: Ilja Turek for his TB-LMTO-SGF layer code; Anton
Starikov for the TB-MTO code based upon sparse matrix techniques.
This work is supported by the National Natural Science Foundation of
China (Grants No. 11104247).

\end{document}